# E-Commerce Delivery Demand Modeling Framework for An Agent-Based Simulation Platform

**Takanori Sakai**
Singapore-MIT Alliance for Research and Technology (SMART)
1 CREATE Way, #09-02 CREATE Tower, Singapore 138602
Email: takanori@smart.mit.edu

**Yusuke Hara**
Next Generation Artificial Intelligence Research Center, The University of Tokyo
7-3-1, Hongo, Bunkyo, Tokyo, Japan, 113-8656
Email: hara@bin.t.u-tokyo.ac.jp

**Ravi Seshadri**
Singapore-MIT Alliance for Research and Technology (SMART)
1 CREATE Way, #09-02 CREATE Tower, Singapore 138602
Email: ravi@smart.mit.edu

**André Alho**
Singapore-MIT Alliance for Research and Technology (SMART)
1 CREATE Way, #09-02 CREATE Tower, Singapore 138602
Email: andre.romano@smart.mit.edu

**Md Sami Hasnine**
Department of Civil and Environmental Engineering, Howard University
2300 Sixth Street, NW #1026, Washington, DC 20059
Email: mdsami.hasnine@howard.edu

**Peiyu Jing**
Intelligent Transportation Systems Lab, Department of Civil and Environmental Engineering,
Massachusetts Institute of Technology
77 Massachusetts Avenue, Room 1-181, Cambridge, M.A., 02139, U.S.A.
Email: peiyu@mit.edu

**ZhiYuan Chua**
NUS Business School, National University of Singapore
1 Business Link, #B2-03 BIZ 2, Singapore 117592
Email: bizzyc@nus.edu.sg

**Moshe Ben-Akiva**
Intelligent Transportation Systems Lab, Department of Civil and Environmental Engineering,
Massachusetts Institute of Technology
77 Massachusetts Avenue, Room 1-181, Cambridge, M.A., 02139, U.S.A.
Email: mba@mit.edu

**ABSTRACT**

The e-commerce delivery demand has grown rapidly in the past two decades and such trend has accelerated tremendously due to the ongoing coronavirus pandemic. Given the situation, the need for predicting e-commerce delivery demand and evaluating relevant logistics solutions is increasing. However, the existing simulation models for e-commerce delivery demand are still limited and do not consider the delivery options and their attributes that shoppers face on e-commerce order placements. We propose a novel modeling framework which jointly predicts the average total value of e-commerce purchase, the purchase amount per transaction, and delivery option choices. The proposed framework can simulate the changes in e-commerce delivery demand attributable to the changes in delivery options. We assume the model parameters based on various sources of relevant information and conduct a demonstrative sensitivity analysis. Furthermore, we have applied the model to the simulation for the Auto-Innovative Prototype city. While the calibration of the model using real-world survey data is required, the result of the analysis highlights the applicability of the proposed framework.

**INTRODUCTION**

E-commerce deliveries has been growing worldwide for the past two decades. In the United States, the share of e-commerce sales for the fourth quarter of 2019 was 11.4%, which was only around 4% in 2009 (U.S. Department of Commerce, 2020). Furthermore, the coronavirus pandemic in 2020 triggered a demand jump for e-commerce deliveries (Fortune, 2020). To cater to the demand increase, a large investment has been made by major e-commerce vendors such as Amazon (The New York Times, 2020). Along with the demand growth, there have been growing discussions among transportation researchers and practitioners on the measures to handle the increasing parcel deliveries to residential locations. Logistics solutions such as last-mile consolidations, collection points/drop zones, cargo cycles, crowdshipping, drones, and delivery robots have been proposed and discussed (e.g., Allen et al., 2018). Furthermore, the relationship between delivery service characteristics and consumers' purchase decisions has been studied from the e-commerce vendors' perspective, including services such as free shipping and same-day deliveries (Nguyen et al., 2019). It should be emphasized that the delivery service options to be offered to e-commerce consumers (called "delivery options" hereafter) rely on transportation network/systems and the both are expected to evolve even more rapidly than before, due to the changes in shopper's preferences toward e-commerce triggered by the pandemic.

Despite its importance for evaluating the impacts of logistics solutions to e-commerce delivery demand, there is a lack of a modeling framework that considers the sensitivity of delivery demand to delivery options, specifically, delivery fee, speed, and time slot, which depends on the performance of transportation systems. Such a framework should also be capable of measuring the impacts of vendor's policies such as free shipping, which is offered typically when the order value exceeds a pre-defined threshold (e.g. $25). The past studies on e-commerce demand models mainly focus on the characteristics of individuals and households but not those of delivery options. This is a critical research gap, as, for example, the delivery fee is known to have significant influence on e-commerce purchases (Lewis et al, 2006; Lepthien and Clement, 2019). Our aim is to develop a modeling framework to answer the question: "how will a consumer react to e-commerce vendor's policy on delivery service options?". In this research, we propose a theoretical demand model framework that predicts e-commerce delivery demand (i.e. frequency, order size, and delivery option choices) given the household characteristics and the offer of delivery options. The proposed framework is expected to be used for simulating e-commerce delivery demand to evaluate the followings from the public policy maker's perspective:

- Impacts of the increase in the purchase share on online shopping against in-store shopping
- Impacts of new delivery options made available by novel transportation modes
- Impacts of the increase/reduction in delivery costs due to transportation system's performance

The fragmentation of commodity flows (i.e. the shift from Business-to-Business (B-to-B) to Business-to-Consumer (B-to-C) flows) is considered to cause more congestion and environmental impacts (Morganti et al., 2014). Thus, the public sector might consider internalizing the externalities by using policy tools such as delivery tax and congestion pricing. The household-based e-commerce demand model is proposed to enhance the urban freight model of an agent-based urban simulation platform named SimMobility (Sakai et al. 2020).

The rest of the paper is organized as follows; a literature review that covers existing e-commerce delivery demand models and the studies on the consumer preferences to delivery options; a description of the proposed modeling framework for e-commerce demand; assumptions on the model parameters, which are set preliminarily, and sensitivity analysis; the demonstrative application of the model to the urban freight simulation as a part of SimMobility; and conclusions.

## LITERATURE REVIEW

There are increasingly more studies that estimate e-commerce delivery demand models, while they are still limited to demand simulation. Here, we focus on the research on demand modeling for city-wide demand simulations. Some studies estimate the models to shed light on the interactions between online and in-store shopping (e.g. Shi et al., 2019; Xi et al., 2020; Suel et al., 2018), which we do not discuss here. A boarder review of online shopping models is available in the literature, for example, Suel and Polak (2018).

As one of the earliest studies focusing on the freight traffic demand estimation, Wang and Zhou (2015) develop a binary choice model and a right-censored negative binomial model to predict delivery frequency, using individual, household, and urban characteristics as independent variables. Their model was estimated using the U.S. National Household Travel Survey (NHTS) data. The studies that follow assume the substitutability between shopping trips and deliveries. Stinson et al. (2019) develop a household-level e-commerce model, which predicts the participation in e-commerce and the ratio of delivery to on-site shopping. They use household characteristics and accessibilities as independent variables. Stinson et al. (2020), using the model in Stinson et al. (2019), estimate parcel delivery truck tours for POLARIS, an agent-based transportation model (Auld et al., 2016). Jaller and Pahwa (2020) develop a multinomial logit model that predicts a shopping decision in each day. Alternatives include "*no shopping*", "*in-store*", "*online*" and "*both*". The 2016 American Time Use Survey (ATUS) was used for model estimation. The independent variables include, gender, age, education level, employment status, family income, mobility-related difficulty, region, the size of the Metropolitan Statistical Area (MSA) and season. Also, Comi and Nuzzolo (2016) propose a probability model considering the four alternatives for weekly purchases similar to those in Jaller and Pahwa. The existing e-commerce demand models for freight traffic demand estimations consider primarily consumer's characteristics, and the offers on delivery options presented by e-commerce vendors are not taken into account.

On the other hand, the consumer's preference for delivery options has been studied mainly in market research. Among them, those estimating choice models relevant to freight demand analysis are below. Regarding the delivery option choice, Nguyen et al. (2019) conduct a conjoint analysis focusing on three product types – a personal care item, a pair of jeans, and a digital camera - and estimate a discrete choice model considering delivery option attributes including delivery speed, time slot, day/evening delivery, delivery date and delivery fee. Gawor and Hoberg (2018) also conduct a choice-based conjoint analysis asking delivery preferences for a digital camera, a laptop, and a smartphone to online shoppers in the US. Similarly, Autry et al. (2011) conduct an adaptive choice-based conjoint analysis, using the data from the students in a university market research class. The attributes such as price, delivery speed, the availabilities of tracking, guaranteed delivery time, insurance, and shipping company are considered. Wilson-Jeanselme and Reynolds (2006) estimate a model with the focus on online grocery shopping.

While several delivery option studies consider the effect of delivery fees among others, the studies on the relationship between the delivery fee and order size are limited. Lewis et al. (2006) is one of a few exceptions. They develop an ordered logit model for order value at online shopping considering the variables such as market and household-level variables and the delivery fee. They use the data from an online retailer specializing in nonperishable grocery and drugstore items.

Furthermore, Lewis (2006) also develop an OLS regression model to predict an average order size (i.e. expenditure) using the same data. Lepthien and Clement (2019) develop another OLS regression model of the order size considering delivery fee. They use data of an online retailer for streetwear and sportswear with the research interest on return behavior. The studies discussed above focus on only a single aspect of e-commerce delivery demand independently from the others. For example, a delivery option selection model does not consider order size decisions. Considering the fact that "free shipping" service is widely offered at present and that the order size is often influential to the delivery fee, those two decisions ("delivery option selection" and "order size selection") are related to each other in the real world.

## PROPOSED MODEL SPECIFICATION

### Overview

We propose a model specification which considers the inter-relations among the decisions on (i) the average total value of e-commerce purchase ("*total value*") per week, (ii) the purchase amount per transaction ("*order value*"), and (iii) delivery option ("*delivery option*"). This is a household-based model and a decision-maker in the model represents her/his household (this "decision maker" could consist of more than one individual online shopper who belongs to a single household). The total value divided by order value gives the delivery frequency. We assume it is critical to consider these three decisions in a joint manner (i.e. a sequential model is not appropriate), because the relationship between delivery options and order and total values is bi-directional:

- The characteristics of delivery options, especially delivery fee, affects the total value and order value; and
- Delivery options' characteristics depend typically on order value, due to free shipping service or other marketing tactics used by e-commerce vendors.

The proposed multi-level model uses a random utility-based estimation framework; the first level is *delivery option*, the second level is *order value*, and the third level is *total value*. On the second and third levels, the expected utility from the lower level is considered through log-term measures (**Figure 1**). It should be noted that the proposed modeling framework does not directly consider weight (i.e. kg) of each order and the potential generation of multiple package deliveries for an order. As for the former, the subsequent use of a value-to-size conversion model is required; as for the latter, the number of packages per order is assumed, while the future enhancement is necessary, taking into account vendors' logistics capabilities (e.g. whether vendors are able to send goods from a single fulfillment center or not).

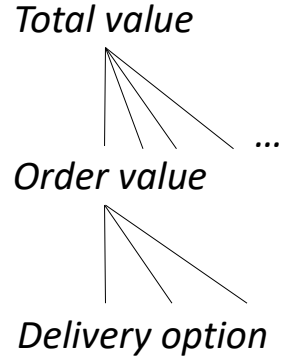

**Figure 1   A multi-level model structure**

## Specification for each level

The model specification of each level is as follows.

### *Delivery Option*

A multinomial logit (MNL) model is used for the delivery option choice. The model considers the situation that a decision maker is solicited to select one option from multiple delivery options (i.e. alternatives) which differ in in terms of speed, slot, time, day, and fee, given the order value. The example of the alternatives is shown in **Table 1.**

**TABLE 1 Example of Delivery Options**

| Option | Speed | Time slot | Time | Date | Fee |
|--------|-------|-----------|------|------|-----|
| 1 | 2-5 days | No time slot | Daytime | Weekday only | US$0 |
| 2 | One day | No time slot | Daytime | Weekday only | US$12 |
| 3 | Same day | 4 hr | Daytime and evening | Weekday and Saturday | US$18 |

Source: Adapted from Nguyen et al. (2019) ("Delivery fee" is different from those presented in Nguyen et al.)

The systematic utility of delivery option ($do$) given order value ($ov$) for a decision maker $n$ is:

$$V_{n,do|ov} = \sum_i \beta_i^{speed} speed_{i,do} + \sum_j \beta_j^{slot} slot_{j,do} + \sum_k \beta_k^{time} time_{k,do} + \sum_l \beta_l^{date} date_{l,do} + \beta_{fee} \log(fee_{do,ov} + 1) \qquad (1)$$

where:

$speed_{i,do}$:   Variable for speed $i$ (1: 2-5 days, 2: one day, 3: same day). 1 if speed category of *do* is $i$; 0 otherwise.

$slot_{j,do}$:   Variable for time slot (width) $j$ (1: no time slot, 2: two hours, 3: four hours). 1 if time slot category of *do* is $j$; 0 otherwise.

$time_{k,do}$:   Variable for time $k$ (1: daytime, 2: daytime & evening). 1 if time category of *do* is $k$; 0 otherwise.

$date_{l,do}$ :  Variable for date $l$ (1: weekday, 2: weekday & Saturday, 3: all days). 1 if date category of $do$ is $l$; 0 otherwise.

$fee_{do,ov}$:  Fee of $do$ given order value $ov$

The set of alternatives $DO_{n|ov} (\ni do)$ is conditional to $ov$ (e.g. free shipping for the standard delivery). Equation (1) is the same with the model specification proposed by Nguyen et al. (2019) except that delivery fee is considered as a continuous variable, instead of a dummy variable.

*Order Value*

A MNL model is used for the order value choice. The set of alternatives includes integer values in US\$, from \$10 to \$300. The systematic utility of order value ($ov$) given total value ($tv$) is:

$$V_{n,ov|tv} = \beta_{logsumdo} \cdot \frac{tv}{ov} \cdot \ln \sum_{do \in DO_{n|ov}} \exp\left(\tilde{V}_{n,do|ov}\right) + \beta_{interval} \cdot \left(\frac{ov}{tv}\right)^2 + \beta_{storage} \cdot ov \quad (2)$$

The first term captures the effect of the total expected utility from the delivery options (frequency per week multiplied by the log-sum measure from the delivery option choice). The second term captures the effect of the interval between orders (i.e. the penalty for long order intervals); it is assumed that the longer interval is associated with the greater depreciation, the mismatch between purchased items and the future consumption need, and the inventory cost. The third term captures the effect of the storage cost, which is linear to the order value.

*Total Value*

Again, a MNL model is used for the total value choice, considering integer values in US\$, from \$1 to \$600 per week, as the set of alternatives. The systematic utility of total value ($tv$) is defined as follows:

$$V_{n,tv} = \beta_{logsumov} \cdot ln \sum_{ov} \exp\left(\hat{V}_{n,ov|tv}\right) + \beta_{hhs} \left(\alpha \cdot hhs_n - tv\right)^2 \quad (3)$$

where:
$hhs_n$  :  Number of people in household $n$.

The first term is the effect of the log-sum measure from the order value choice. The second term captures the "gap" between the household size; the model parameter $\alpha$ indicates the scale of per-household member e-commerce purchase need.

The modeling framework consisting of the above three levels allows the e-commerce delivery demand prediction to be sensitive to delivery options' attributes on multiple dimensions, i.e. frequency, size, and delivery requirements (e.g. standard or express delivery). Ideally, the model parameters would be calibrated for the three levels simultaneously; however, there is no real-world data available for us to estimate model parameters altogether. It must be noted that the data collection effort is on-going by the authors to collect the online and offline shopping information jointly with the daily activity records using the enhanced Future Mobility Sensing (FMS) (Zhao et al., 2015), called FMS E-Commerce.

**ASSUMPTIONS ON MODEL PARAMETERS AND SENSITIVITY ANALYSIS**

In this section, we assume model parameters based on the past studies and/or synthetic data (instead of using the real-world data for the full model estimation) and use them for the demonstrative sensitivity analysis. While we attempt to use realistic model parameters, we do not claim the model parameters and simulation results represent any real-world circumstances. Our claimed research contribution is conceptual only and this section is to demonstrate the practicality of the proposed framework. Further calibration is required for using the models for predicting the real-world e-commerce delivery demand.

**Assumptions**

Model parameters are set based on information from various sources. The summary of assumed model parameters is shown in **Table 2**. The belief explanation of the parameter setting for each level is as follows:

**Delivery Option**: The parameters are taken from the estimated model in Nguyen et al. (2019). They conducted a conjoint analysis to estimate the part-worth utilities for the delivery option choice. Their analysis used the data from 1012 respondents in the Netherlands, who expressed delivery option preferences for the purchase of *personal care item* (representing *Convenience Goods*), *a pair of jeans* (representing *Shopping Goods*), and *a digital camera* (representing *Specialty Goods*). We use the parameters estimated for *personal care item* while we make changes on the delivery fee term. In the original model, the delivery fee is treated as a categorical variable (i.e. €0, €2.5, €4, €7.5 or €17.5). We replace this with a function that treats the delivery fee as a continuous variable as shown in Equation (1); the coefficient for the delivery fee is calibrated to replicate the part-worth utility in the original model and the unit of currency is converted from € to US$.

**Order Value:** The available data/models which capture the relationship between the order value and delivery option characteristics, especially the delivery fee, are limited. We rely on Lewis et al. (2006) to generate synthetic data for parameter estimation. Lewis et al. (2006) estimated an ordered logit model to predict order size, using the data from an online retailer specializing in *non-perishable grocery and drugstore items*. The data collection year and location are not disclosed but supposedly the data is from the US in 2003 or earlier. We generate the data using their model, which consider the relationship between delivery fee and order size. To generate the data, we assume four fee structures for small ($0-$50), medium ($50-$75) and large ($75-) order size (i.e. $5/$7/$0, $5/$7/$9, $3/$5/$5, and $6/$8/$10). Since the responses by Lewis et al. (2006) model are categorical ($0-$50, $50-$75, $75-), randomly selected continuous values are assigned instead of categorical values. To estimate the model parameters for Equation (2), we use the model parameters of the Delivery Option model shown in Table 2.

**Total Value:** We use the household survey from 2019 Puget Sound Regional Travel Study (RSG, 2020). The study collected the data of no. of packages delivered in a day for each household member for a week. We compute the number of package deliveries for the survey week for each household and, assuming each delivery order generates three packages on average (this assumption leads to 0.5 order/week/adult on average; Amazon prime members placed 24 orders per year in 2018 (Statista, 2020a)), generate the order frequency data. Then, we assign order values to each household using the order value – frequency bivariate distribution from the aforementioned synthetic data for the Order Value model estimation, compute the total value for each household, and further adjust these values by multiplying a factor to make an average household-level total demand $50 per week (this is set referring to the data of Amazon Prime members; average spend by an Amazon Prime member (i.e. individual) in the US is $1.4 thousand per year and $27 per week (Statista, 2020b)).

To estimate the model parameters for the Total Value model (Equation (3)), we use the parameters of the Delivery Option and Order Value models estimated earlier, as shown in Table 2. It is also assumed that households face the three delivery options with varying delivery fees by order size shown in **Table 3**.

**TABLE 2 Summary of Assumed Model Parameters**

| Parameter | Value | Parameter | Value |
|---|---|---|---|
| **Delivery option choice** | | **Order value choice** | |
| $\beta_i^{speed}$ | | $\beta_{logsumdo}$ | 1.05 |
| - same day | 0.177 | $\beta_{interval}$ | -0.111 |
| - next day | 0.082 | $\beta_{storage}$ | -0.0183 |
| - 2-5 business days | -0.259 | **Total value choice** | |
| $\beta_j^{slot}$ | | $\beta_{logsumov}$ | 0.0597 |
| - no time slot | -0.157 | $\beta_{hhs}$ | -0.000175 |
| - 2 hr | 0.113 | $\alpha$ | 12.3 |
| - 4 hr | 0.040 | | |
| $\beta_k^{time}$ | | | |
| - daytime | -0.090 | | |
| - daytime & evening | 0.090 | | |
| $\beta_l^{date}$ | | | |
| - weekday | -0.063 | | |
| - weekday & Saturday | 0.054 | | |
| - all days | 0.009 | | |
| $\beta_{fee}$(fee, US$) | -1.377 | | |

**TABLE 3 Assumed Delivery Options for the Total Value Model Estimation**

| Delivery option (*do*) | Delivery Speed | Delivery fee Order size: | | | | Other characteristics |
|---|---|---|---|---|---|---|
| | | $0-25 | $25-50 | $50-100 | $100- | (same for all options) |
| 1 | 2-5 days | 6 | 3 | 3 | 3 | No time slot, daytime |
| 2 | One day | 12 | 15 | 17 | 20 | delivery, and all days |
| 3 | Same day | 18 | 20 | 22 | 27 | |

**Sensitivity Analysis**

Using the assumed model parameters listed in **Table 2**, we conduct a sensitivity analysis. As mentioned earlier, this analysis is for the demonstrative purpose only, highlighting the way the proposed model specification makes the demand characteristics, i.e. e-commerce order size, order frequency and delivery option choice, sensitive to the delivery options, specifically delivery fees, offered by e-commerce vendors. We use the household size data generated based on the household data from 2019 Puget Sound Regional Travel Study (933 households that received packages at least once during the survey period). It should be noted that the information used is only the household size (i.e., no. of individuals in each household), which is an independent variable in Equation 3.

We set four delivery option scenarios as shown in **Table 4**. In each scenario, the same set of delivery options is assumed for all households. S1 is the scenario with free shipping service; when the order size exceeds $25, the delivery fee of DO1 (2-5 days) is zero. S2 is the scenario without such a free shipping service; a shopper needs to pay delivery fee even if the order size is greater than $25. S3 and S4 are also the scenarios with and without free shipping service, respectively. The difference between S1 & S2 and S3 & S4 are the delivery fee for the express deliveries (i.e. one day and same day deliveries); 30% less express delivery fee is assumed for S3 & S4, compared with S1 & S2.

**TABLE 4 Delivery Option Scenarios**

| Scenario | Delivery option (do) | Speed | Fee (US$) Order size: | | | |
|---|---|---|---|---|---|---|
| | | | $0-25 | $25-50 | $50-100 | $100- |
| **S1**: Free shipping | 1 | 2-5 days | 6 | 0 | 0 | 0 |
| | 2 | One day | 12 | 15 | 17 | 20 |
| | 3 | Same day | 18 | 20 | 22 | 27 |
| **S2**: No free shipping | 1 | 2-5 days | 6 | 7 | 8 | 10 |
| | 2 | One day | 12 | 15 | 17 | 20 |
| | 3 | Same day | 18 | 20 | 22 | 27 |
| **S3**: Free shipping & 70% delivery fee for express deliveries | 1 | 2-5 days | 6 | 0 | 0 | 0 |
| | 2 | One day | 8.4 | 10.5 | 11.9 | 14 |
| | 3 | Same day | 12.6 | 14.0 | 15.4 | 18.9 |
| **S4:** No free shipping & 70% delivery fee for express deliveries | 1 | 2-5 days | 6 | 7 | 8 | 10 |
| | 2 | One day | 8.4 | 10.5 | 11.9 | 14 |
| | 3 | Same day | 12.6 | 14.0 | 15.4 | 18.9 |

Note: The other characteristics are the same across scenarios and delivery options: No time slot, daytime delivery, and all days (i.e. both weekday & weekend).

The simulation result is shown in **Figure 2 and 3** and **Table 5**. **Figure 2** shows the cumulative density distribution of total value per week. The differences between S1 and S3 and between S2 and S4 are limited, indicating that the effect of the express delivery fee reduction is limited in the assumed setting. A clearer difference is observed between S1 and S2 (or S3 and S4). The suspension of free shipping service leads to about 6% reduction in the total value. **Figure 5** shows the cumulative density distribution of the order value. The figure highlights a remarkable difference between the scenarios with free shipping and those without free shipping; delivery frequency is about 36% less in S2 compared with S1. The model describes shoppers' preference to reduce the delivery fee by increasing order size and decreasing order frequency.

**Table 5** summarizes the result including the share of deliveries by delivery speed. The result shows that, when comparing the scenarios with and without free delivery service, the reduction of 2-5 days delivery is much greater when express delivery fee is cheaper. (The difference between S3 and S4 is 50.2% while that between S1 and S2 is 42.5%.) While the plausibility of these simulated sensitivities is still required to be examined, the analysis shows the modeling framework can describe the intended demand sensitivities.

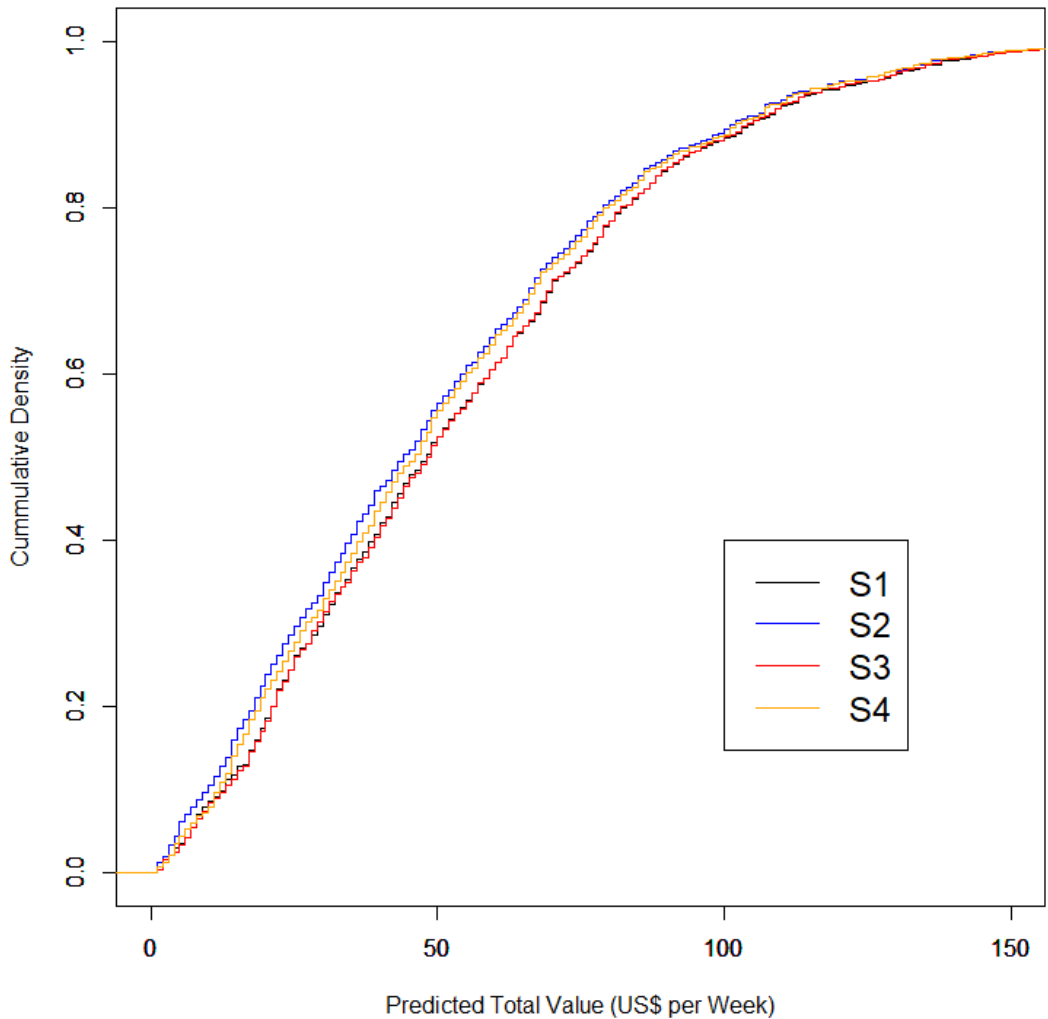

**Figure 2   Comparison of total values**

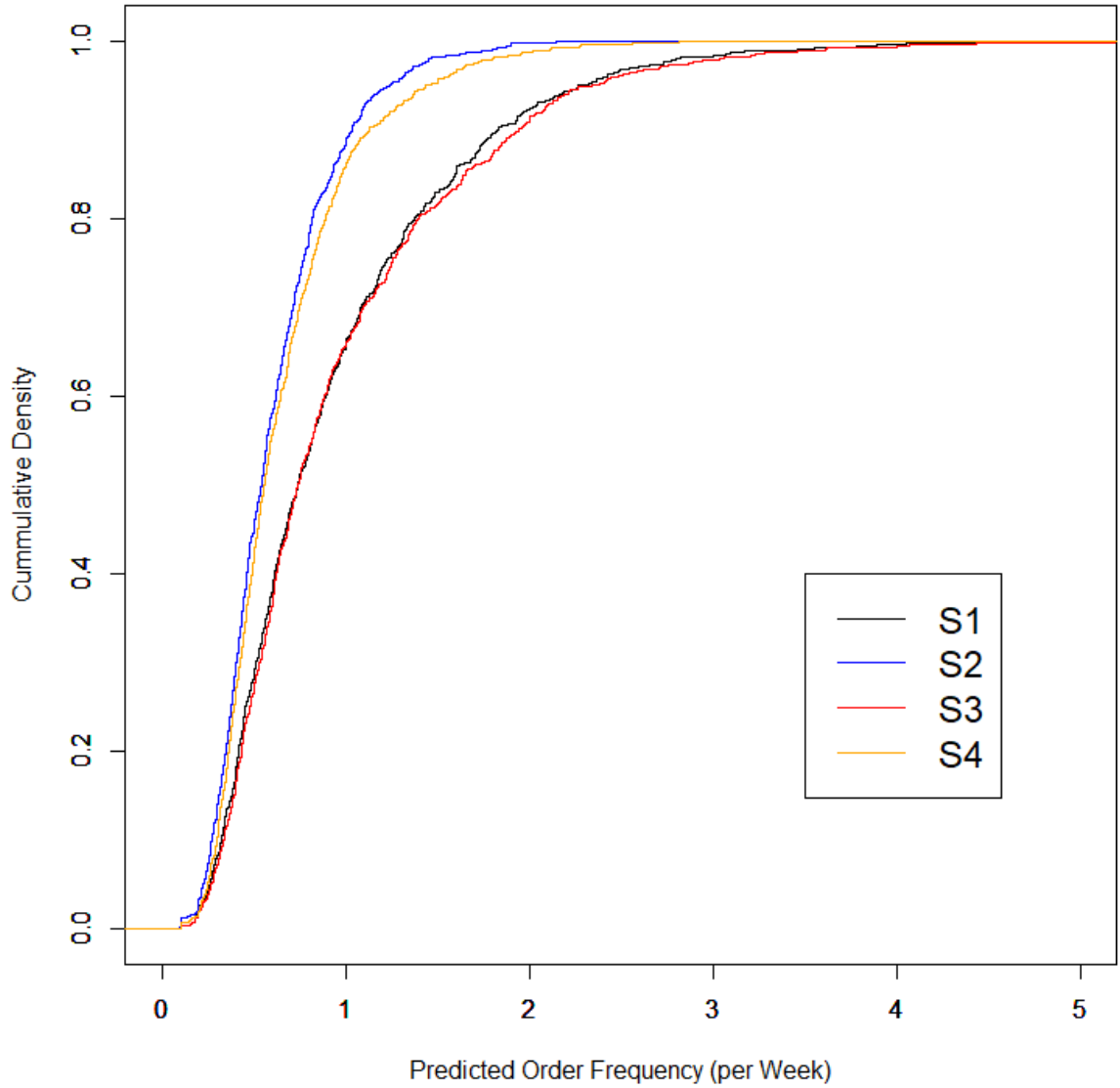

**Figure 3   Comparison of order frequencies**

**TABLE 5 Summary Statistics of Sensitivity Analysis**

|  | Total Value: Mean | Order Frequency: Mean | Share of 2-5 days delivery (%) | Share of one day delivery (%) | Share of same day delivery (%) |
|---|---|---|---|---|---|
| S1 | 54.2 | 0.944 | 93.3 | 3.9 | 2.8 |
| S2 | 50.8 | 0.607 | 50.8 | 28.1 | 21.1 |
| S3 | 54.2 | 0.971 | 89.8 | 5.9 | 4.3 |
| S4 | 52.0 | 0.659 | 39.6 | 34.3 | 26.0 |

## APPLICATION FOR AUTO-INNOVATIVE CITY

Despite the rapid increase in e-commerce deliveries in the recent years, the significance of B-to-C e-commerce deliveries to urban traffic, compared against other freight traffic, is not well-known. One of the reasons for this is that the widely used surveys for classified traffic counts cannot differentiate goods vehicles for e-commerce deliveries and those for other purposes. As another demonstrative analysis, we use SimMobility Freight enhanced with the proposed e-commerce model, which are calibrated based on the available statistics, to obtain the estimation of e-commerce traffic in a metropolitan area. We use the Auto-Innovative prototype city (Boston as an archetype) for this analysis. The prototype city is a model of a class (or typology) of cities. The Auto-Innovative Prototype represents a dense auto-dependent North American city with high transit mode share and population density (Oke et al., 2019; Oke et al., 2020). The synthetic household and establishment data developed for the Auto-Innovative prototype city are used as inputs; the establishments also include distribution facilities for e-commerce deliveries.

### Simulation Flow

The flow of the e-commerce demand simulation is shown in Figure 4. The households is the main input for the simulation. The e-commerce adoption rates were set as 0.12, 0.30 and 0.50 for each of groceries, household goods & medicines, and other packages, considering the information from various sources (for example, Etumnu and Widmar (2020)). The "adoption" is defined here the placement of at least one order in a random month. These adoption rates were finalized after the calibration process described later. The households were randomly selected based on the adoption rate for each item category and the household-based e-commerce demand model is applied to the selected households. The e-commerce orders are converted to packages and, for each e-commerce package, a distribution facility is assigned using a discrete choice model (the model used in Supplier Selection Module of SimMobility Freight (Sakai et al., 2020)). The output, e-commerce shipments, was put together with B-to-B shipments, and used as the input for Pre-day Logistics Planning in SimMobility Freight.

### Calibration

The key information on the model calibration for B-to-B shipments, e-commerce shipments, and the associated vehicle operation planning are as follows:

*B-to-B shipments:* The initial model parameters were estimated using the data from 2013 Tokyo Metropolitan Freight Survey, which has large sample size and detailed information of shipments, including shipper and receiver information. The detailed information of the data and parameters is available in Sakai et al. (2020). We re-calibrated the model parameters based on the 2012 Commodity Flow Survey – Public Use Microdata (2012 PUM) (U.S. Census Bureau, 2012). The shipment records with shipper industry type and commodity type information are available from 2012 PUM but the data is mainly for manufacturers and wholesalers; therefore, we used Freight Analysis Framework (FAF4) data, which provide aggregate OD flow, to complement 2012 PUM. Models were calibrated so that aggregate freight generation matches the survey data. Furthermore, shipment size model parameters were calibrated for shipments from manufacturing and wholesale industries for each commodity type.

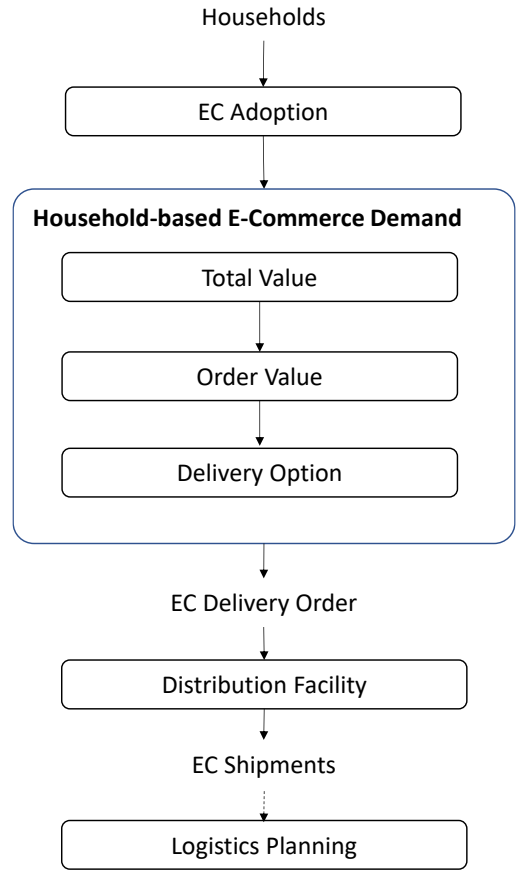

**Figure 4   Flow of E-commerce demand simulation**

*E-commerce orders*: Initial model parameters are those described in the previous section. We further adjusted the parameters, mainly those relevant to $\alpha$ in Equation (3) (*the scales of per-HH member e-commerce purchase need*, $\beta_{interval}$ and $\beta_{storage}$ in Equation (2) for each of item categories. Such adjustment was made so that the number of deliveries and average monthly e-commerce expenditures match the reference values computed based on available statistics (Holguin-Veras and Wang, 2020; The Atlantic, 2019; U.S. Bureau of Labor Statistics; 2019). The figures used as reference are shown in **Table 6**.

**TABLE 6   Reference for Calibration**

| Reference for calibration | Groceries | Household goods and medicines | Other packages |
|---|---|---|---|
| No of deliveries (pre-Pandemic)[a] | 0.6/person-month *(exclude <15 yrs old)* | 1.2/person-month *(exclude <15 yrs old)* | 3.1/person-month *(exclude <15 yrs old)* |
| Ave. total monthly purchase value [b] | 11 USD/HH (3% of total expense) | 38 USD/HH (26% of total expense) [c] | 62 USD/HH (26% of total expense) [c] |

Note: a) Based on Holguin-Veras & Wang (2020); b) Assumed based on The Atlantic (2019) and U.S. Bureau of Labor Statistics (2020).

*Vehicle operation planning*: Various parameters were set in the vehicle operations planning module. The two key settings are as follows:

o   Dwell time: Average dwell time per parcel was set as 2.3 mins. (Allen et al. (2018) report that the mean parking time per parcel is 2.3 minutes based on a survey in London.)

o   Vehicle capacity: The vehicle volume capacity and the weight-to-volume conversion is set so that average number of parcel deliveries per tour becomes around 18 for groceries and 44 for non-grocery items.

**Result**

Figure 5 shows the number of deliveries per driver-day. As a rough indication, Kuo (2018) reported that an average FedEx driver can delivery around 75-125 packages per day while the average of the home delivery industry is 15-35 packages per day (Jaller and Pahwa, 2020). The result is more or less consistent with this anecdotal evidence.

Figure 6 shows the number of vehicles under operation by time of day. The contribution of e-commerce delivery vehicles is significantly high; about 20-30% of goods vehicles under operation during the daytime is those for e-commerce deliveries. While the validation using the traffic count data is required in the future, this result, which is calibrated at the shipment level using the available data, indicates that e-commerce-driven traffic demand has grown to the level which is no longer negligible. It must be noted that the household-based e-commerce demand model was calibrated against the pre-Covid-19 pandemic standard.

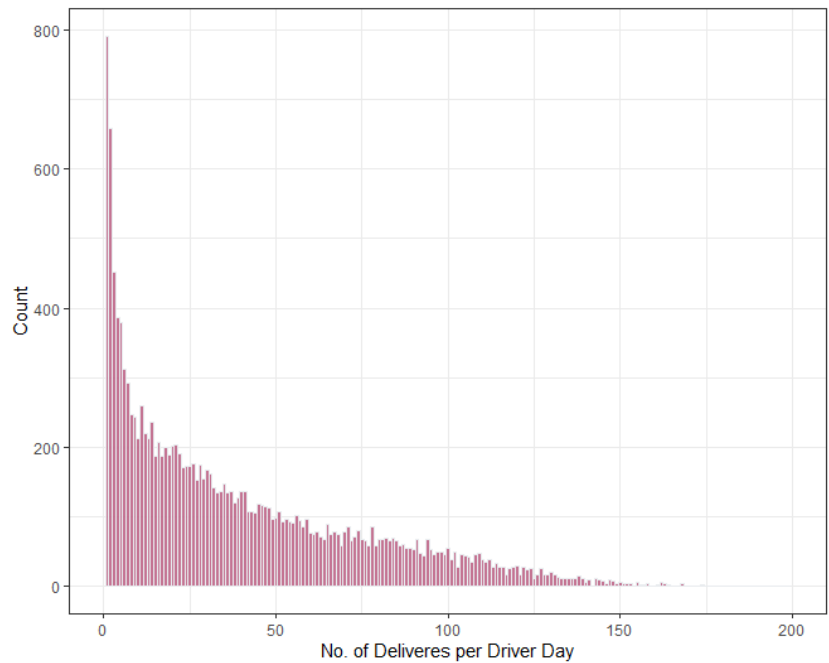

**Figure 5   The number of deliveries per driver day (E-commerce deliveries)**

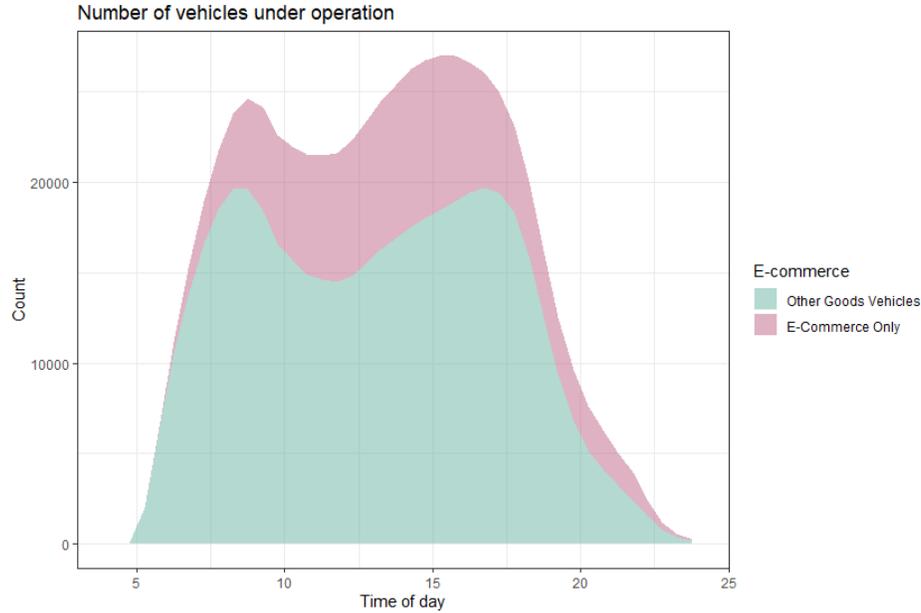

**Figure 6   The number of vehicles under operation**

## CONCLUSIONS

This research proposes an innovative modeling framework for e-commerce delivery demand. The proposed framework jointly predicts the average total value of e-commerce purchase, the purchase amount per transaction, and delivery option choices. In the framework, the demand is sensitive to delivery option attributes, which is of critical importance to evaluate the impacts of novel urban logistics solutions that affect delivery speeds and costs. Using the assumed model parameters, we demonstrate how the model performs. In the simulations using the model, the changes in delivery fee for different delivery speeds affect the three dimensions of e-commerce delivery demand. We also integrate the model to SimMobility and applied to the traffic analysis for Auto-Innovative prototype city.

Clearly, this research contributes only to the first step in developing a comprehensive modeling framework of the e-commerce delivery simulation. The expected next step of the work presented in this paper includes the full calibration of the model using real-world data. As for the proposed framework, there is room for incorporating heterogeneity attributable to household characteristics, commodity types and the different order situations (e.g. a shopper who requires some specific item immediately vs. a shopper who places recurring orders). Furthermore, developments on supply-side simulations are required to rigorously evaluate the impacts of supply improvement. Moreover, the interaction between e-commerce and in-store shopping is, as identified in other studies (Suel and Polak, 2018), important for comprehensively evaluating the traffic impacts of the growth in e-commerce.

## ACKNOWLEDGEMENTS

This research is supported in part by the National Research Foundation, Prime Minister's Office, Singapore, under its CREATE programme, Singapore-MIT Alliance for Research and Technology (SMART) Future Urban Mobility (FM) IRG. It is also supported in part by the Singapore Ministry of National Development and the National Research Foundation, Prime Minister's Office under the Land and Liveability National Innovation Challenge (L2 NIC) Research Programme (L2 NIC

Award No. L2NICTDF1-2016-1). Any opinions, findings, and conclusions or recommendations expressed in this material are those of the authors only.

**AUTHOR CONTRIBUTIONS**